\documentclass[reprint,superscriptaddress,amsmath,amssymb,aps,floatfix,prl]{revtex4-2}
\usepackage{amsmath}
\usepackage{graphicx} % Include figure files
\usepackage{dcolumn}  % Align table columns on decimal point
\usepackage{bm}       % bold math
\usepackage{xcolor}   % Color the text.
\usepackage{hyperref} % add hypertext capabilities
\usepackage{upgreek}
\usepackage{comment}

\begin{document}
\title{Detuning-symmetric laser cooling of many mechanical modes with a photothermally modified cavity}

\author{Thomas J. Clark}
\email{tommy.clark@mail.mcgill.ca}
\author{Jiaxing Ma}
\author{Jack Sankey}
\email{jack.sankey@mcgill.ca}
\affiliation{McGill University Department of Physics}
\date{\today}

%%%%%%% I say effective mass a lot. Let's make some macros 
% These have the $$ in them so they can be used in Text environments direcrly
\newcommand{\me}{$m_{\mathrm{eff}} \ $}
\newcommand{\Om}{$\Omega_{\mathrm{m}} \ $}
\newcommand{\Gm}{$\Gamma_{\mathrm{m}}\ $}
\newcommand{\tm}{$\tau_{\mathrm{m}}\ $}

\newcommand{\?}[1]{\textcolor{red}{[[ #1 ]]}}
\newcommand{\nb}[1]{\textcolor{blue}{[[ #1 ]]}}

\newcommand{\TE}{\mathcal{T}_\text{E}}
\newcommand{\TR}{\mathcal{T}_\text{R}}
\newcommand{\SiO}{SiO$_2$}
\newcommand{\TaO}{Ta$_2$O$_5$}

% ABSTRACT %%%%%%%%%%%%%%%%%%%%%%%%%%%%%%%
%
\begin{abstract}
We simultaneously cool $\gtrsim$100 mechanical modes of a membrane with a photothermally modified optical cavity driven by a single blue-detuned laser. 
In contrast to radiation pressure and bolometric forces applied directly to the mechanical system, this cooling effect does not depend on the sign of detuning, allowing for single-laser stabilization (i.e., simultaneous positive optical spring and damping) that is especially effective at room temperature and high laser power.
We also provide intuition about the competing thermal processes, and propose two simple modifications to the mirror coatings that can strongly enhance this effect.
\end{abstract}
\maketitle

\textit{Introduction.---} 
As cavity optomechanical systems \cite{Aspelmeyer2014Dec} at room temperature continue to improve sensitivity -- even approaching the limits imposed by quantum mechanics \cite{Aasi2013Aug, Cripe2019Apr, Aggarwal2020Jul} -- it is essential to develop robust stabilization techniques to minimize the thermomechanical noise \cite{saulson1990thermal}, including (for the most sensitive systems) thermal intermodulation noise \cite{Fedorov2020Nov,Pluchar2023Nov}). This is particularly important in phononic crystal membranes, known for producing high-quality ``defect'' modes with quantum operation \cite{Tsaturyan2017Aug, Mason2019Aug} at room temperature \cite{Huang2024Feb}, since their dense population of mechanical modes \textit{outside} the band gap produce excessive detuning noise -- so much so that multimode active feedback is required to stabilize these systems even at low temperatures \cite{Rossi2018Nov}.
In the commonplace ``fast-cavity limit" (wherein the mechanical frequency is slower than the cavity decay rate), a traditional optical spring \cite{Cuthbertson1996Jul, Sheard2004May, Arcizet2006Nov} can provide some stabilization through increased mechanical stiffness, but since radiation forces normally lag relative to mechanical motion, this is accompanied by a \emph{destabilizing} negative optical damping. 
By the same token, a stabilizing positive optical damping (laser cooling) is is necessarily accompanied by a destabilizing negative spring \cite{Aspelmeyer2014Dec}. This limitation can be mitigated by a second laser \cite{corbitt2007all} or second cavity mode \cite{Singh2016Nov}, exploiting differences in the functional dependencies of the optical spring and damping on detuning. In both cases, however, the competition between the two effects necessitates significantly higher power (in addition to added technical overhead).
Another approach is to utilize active feedback, by either directly providing optical damping \cite{Corbitt2007Oct} or by feeding back to the cavity length \cite{mow2008cooling} or laser frequency \cite{Schediwy2008Jan} to modify the effective cavity susceptibility, leading to enhanced optical stabilization. However, active feedback necessarily introduces loop delay and additional technical challenges that practically limits its scalability to many modes. 

Recent studies have suggested \cite{ballmer2015photothermal, kelley2015observation} and demonstrated \cite{altin2017robust} that mirror materials can exhibit photothermal effects that, with essentially no loop delay, dynamically modify the effective cavity length in response to cavity power, leading to enhanced optomechanical damping that remains positive regardless of whether the drive laser is blue or red detuned from resonance. 
Thus far these effects have been observed with a single low frequency ($\ll$1 kHz) mechanical mode, wherein positive damping with a blue-detuned laser is achieved only by incorporating a thick glass layer inside the cavity (to boost the thermorefractive effects, as discussed below), while red detuning destabilizes the system due to excessive anti-spring. As noted in Refs.~\cite{ballmer2015photothermal, kelley2015observation}, standard mirror coatings can in principle produce stable traps for sufficiently high mechanical frequencies, but this regime has yet to be realized. 

Here we demonstrate that a high-finesse membrane-fiber-cavity system can simultaneously generate positive optical spring and damping, thereby stabilizing the motion of the lowest $\gtrsim$100 mechanical modes. The added optical damping is notably positive for both signs of detuning, and follows the expected detuning and power dependence associated with a photothermally modified cavity \cite{altin2017robust}. We conclude by suggesting two simple modifications to the mirror coatings that should enhance this stabilization, namely (i) partially etching the high-index dielectric layer \cite{Bernard2020Nov} and (ii) embedding a \SiO~Fabry-Perot spacer within the coating. This stabilization technique is well-suited to room-temperature systems at high laser power, especially in situations where traditional (red-detuned) cooling is destabilized by the associated negative optical spring.

\begin{figure}[t!]
\centering
  \includegraphics[width=0.47\textwidth]{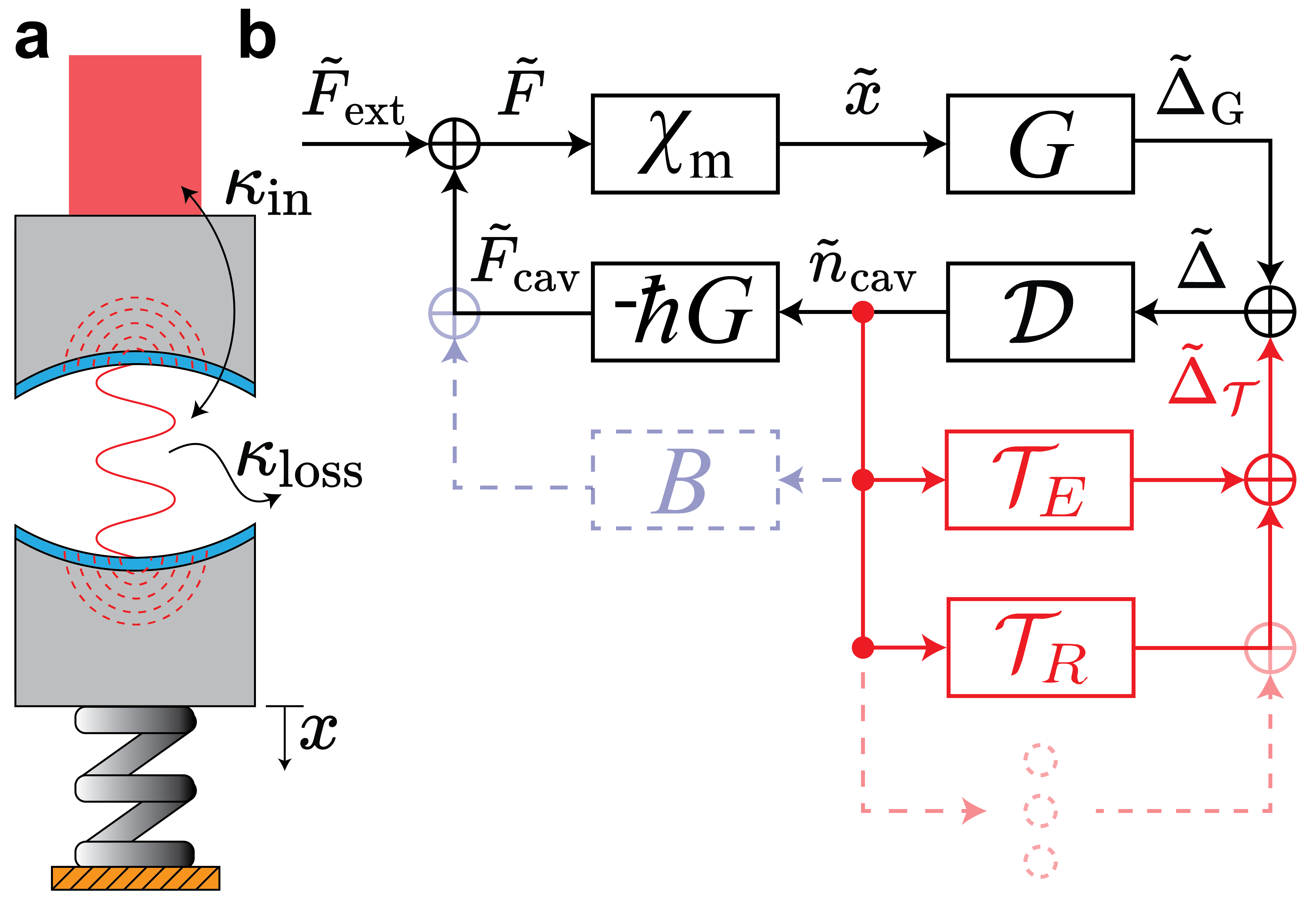}
  \caption{ \label{fig1} 
  Canonical optomechanical system with photothermal effects.
  a) Schematic. Two mirrors form an optical cavity with total energy decay rate $\kappa =\kappa_{\rm{in}}+\kappa_{\rm{loss}}$.
  The displacement $x$ of the bottom mirror, a mass-spring harmonic oscillator, modulates the cavity length and detuning $\Delta$ between the cavity resonance (red standing wave) and input laser (top red block) frequencies. Dotted red lines represent heat from absorbed cavity light, which, through thermorefractive and thermoelastic effects, also modulates $\Delta$.  
  b) Feedback loop describing the system, including radiation pressure (black) bolometric forces (blue) and thermal effects in the mirrors (red). For radiation pressure, the nominal mechanical susceptibility ($\chi_{\rm m}$) is modified by a feedback loop comprising optomechanical coupling $G$, the cavity's susceptibility $\mathcal{D}$ to detuning  fluctuations $\tilde{\Delta}$ (about mean value $\bar{\Delta}$), and radiation pressure ($-\hbar G $). The red loops describes photothermal mechanisms that modify the bare cavity transfer function, producing the additional stabilization reported here. Here these are dominated by thermoelastic ($\mathcal{T}_\text{E}$) and the thermo-refractive ($\mathcal{T}_\text{R}$) effects.}
\end{figure}

\textit{Photothermal Model.---}
As discussed in Ref.~\cite{altin2017robust, Metzger2004Dec}, photothermal effects offer the possibility of improving mechanical stability via optical spring and optical damping simultaneously.
Figure 1(a) shows a schematic of the canonical optomechanical system \cite{Aspelmeyer2014Dec} comprising an optical cavity with one movable mirror acting as a harmonic oscillator of mass $m$, resonant frequency $\Omega_\text{m}$, damping rate $\Gamma_\text{m}$, and bare mechanical susceptibility
\begin{equation}
    \chi_\text{m}=\frac{1/m}{\Omega_\text{m}^{2}-\omega^{2}-i\Gamma_\text{m}\omega}.
\end{equation} 
The cavity's energy decay rate $\kappa=\kappa_\text{in}+\kappa_\text{loss}$ comprises the input mirror's transmission ($\kappa_\text{in}$) and all other losses ($\kappa_\text{loss}$, including absorption). To populate the cavity, the input mirror is driven by $p_\text{in}$ photons per second, detuned by $\Delta$ from the cavity's resonance, and the resulting $n_\text{cav}$ cavity photons both influence and are influenced by the mirror's displacement $x$; this can generate an optical spring and damping that modify $\chi_\text{m}$ \cite{Aspelmeyer2014Dec}.

The effects of radiation pressure are well-described by the black feedback loop in Fig.~\ref{fig1}(b). To understand how $\chi_\text{m}$ changes, we examine how an external oscillatory force $\tilde{F}_\text{ext}e^{i\omega t}$ induces a commensurate displacement $\tilde{x}e^{i\omega t}$ (about equilibrium $\bar{x}$). 
With the loop closed, the \textit{total} force amplitude $\tilde{F}=\tilde{F}_\text{ext}+\tilde{F}_\text{cav}$  drives displacement $\tilde{x}=\chi_\text{m}\tilde{F}$, which detunes the cavity by $\tilde{\Delta}=G\tilde{x}$ via the optomechanical coupling $G$. This in turn modulates the photon number by $\tilde{n}_\text{cav} = \mathcal{D}\tilde{\Delta}$, where
\begin{equation}\label{eq:C}
\mathcal{D} =
\left(\frac{ \kappa_{\rm in}p_\text{in}} {\bar{\Delta}^{2}+\frac{\kappa^2}{4}}\right)
\left( 
\frac{1}{\bar{\Delta}+\omega + i \frac{\kappa}{2}}
+
\frac{1}{\bar{\Delta}-\omega -  i \frac{\kappa}{2}}\right),
\end{equation} 
is the cavity's susceptibility to a modulated \textit{detuning} (about equilibrium $\bar{\Delta}$). Finally, the cavity photons apply a modulated radiation-pressure force on the movable mirror with amplitude $\tilde{F}_\text{cav}=-\hbar G \tilde{n}_\text{cav}$ (about equilibrium $\bar{F}_\text{cav}=-\hbar G \bar{n}_\text{cav}$). 
Solving the resulting loop equation $\tilde{x} = \chi_\text{m}\tilde{F} = \chi_\text{m}\left( \tilde{F}_\text{ext} - \hbar G^2C \tilde{x} \right)$ for $\tilde{x}$ yields a loop-modified mechanical susceptibility to $\tilde{F}_\text{ext}$ having resonant frequency $\Omega_\text{eff}=\sqrt{\Omega_\text{m}^2+\Omega_G^2}$ shifted by optical spring constant
\begin{equation}\label{eq:Kopt} %Single # and label for split equation
   \begin{aligned}
   m \Omega_G^2 &\approx 
   \hbar G^{2}\text{Re}\left[\mathcal{D}\right]  \\ 
    & = 2 \hbar G^2 \kappa_{\rm in} 
   \frac{ \bar{\Delta}}{(\kappa^2 / 4 + \bar{\Delta}^2)^2}p_\text{in} 
   \end{aligned} 
\end{equation}
and a damping rate $\Gamma_\text{eff} = \Gamma_\text{m}+\Gamma_G$ optically adjusted by
\begin{equation}
\begin{aligned}\label{eq:Gamma_opt}
    \Gamma_G &\approx - \frac{\hbar G^{2}}{m\Omega_\text{m}}\text{Im}\left[\mathcal{D}\right] \\
    &= \frac{-2 \hbar G^2 \kappa \kappa_{\rm in}}{m}  
   \frac{ \bar{\Delta}}{(\kappa^2 / 4 + \bar{\Delta}^2)^3}p_\text{in} .
\end{aligned}
\end{equation}
Importantly, the real (imaginary) part of the cavity's detuning susceptibility $\mathcal{D}$ determines the optical spring (damping), and both are \textit{antisymmetric} functions of detuning $\bar{\Delta}$. As such, in the fast-cavity limit ($\kappa>\omega$), stabilizing optical spring (damping) is always accompanied by a \textit{destabilizing} antidamping (antispring) \cite{Aspelmeyer2014Dec}, placing upper bounds on the achievable strength of either. 

The influence of photothermal effects in the mirrors are well-described by the red loop, which has a transfer function 
\begin{equation}\label{eq:T}
\mathcal{T}=\mathcal{T}_\text{E}+\mathcal{T}_\text{R}
\end{equation}
arising from thermoelastic ($\mathcal{T}_\text{E}$) and thermorefractive ($\mathcal{T}_\text{R}$) effects, both of which convert cavity photon modulation $\tilde{n}_\text{cav}$ directly to detuning modulation $\tilde{\Delta}_\mathcal{T}$ by adjusting the effective cavity length. 
Solving the \emph{lower} loop equation $\tilde{n}_\text{cav}=\tilde{\Delta}\mathcal{D}$ for $n_\text{cav}$ alone (now with $\tilde{\Delta}_\text{m}$ playing the role of ``external drive'') similarly yields an effective cavity detuning susceptibility
\begin{equation}
    \mathcal{D}_\text{eff} = \frac{\mathcal{D}}{1-\mathcal{T}\mathcal{D}} 
    \approx \mathcal{D}+\mathcal{T}\mathcal{D}^2
\end{equation}
when the effect is sufficiently small (e.g., at low power $p_\text{in}$ or absorption, when heating is small). The last expression provides quick intuition: In the fast-cavity limit ($\kappa\gg \omega$), $\tilde{n}_\text{cav}$ responds nearly adiabatically to $\tilde{\Delta}$, and so $\mathcal{D}$ is mostly real-valued. On the other hand, if thermal time scales are \emph{slow} compared to mechanical dynamics, the correction term $\mathcal{T}\mathcal{D}^2$ can be largely \textit{imaginary}, providing additional damping or antidamping proportional to $\mathcal{D}^2 \propto p_\text{in}^2$ (see Eq.~\ref{eq:C}), depending on the sign of $\text{Im}[\mathcal{T}]$. Other effects (e.g., dielectric nonlinearities) can also be included in this loop, but these should have a comparatively small effect on detuning, and are anyway too fast to generate a significant imaginary $\mathcal{D}_\text{eff}$. In the fast-cavity limit, photothermal effects lead to an additional damping
\begin{align}\label{eq:Gamma_pt}
\Gamma_\mathcal{T} &\approx 
\text{Im} [\mathcal{T}] 
\frac{4 \hbar G^2 \kappa_{\rm in}^2 }{\Omega_{\rm m} m}
\frac{\bar{\Delta}^2} {\left(  \kappa^2 / 4 + \bar{\Delta}^2 \right)^{4}}p_{\text{in}}^{2}.
\end{align}
Notably, $\Gamma_\mathcal{T}$ is a \textit{symmetric} function of detuning, providing access to simultaneous stability of the optical spring and damping at sufficiently high power. 

In practice, thermoelastic effects ($\mathcal{T}_\text{E}$) tend to grow mirrors inward and \textit{shorten} the cavity when power is absorbed, which generates antidamping. However, thermorefractive effects ($\TR$) tend to \textit{lengthen} the cavity by increasing the index of refraction in the coatings, thereby generating damping \cite{altin2017robust}. This competition can produce net damping or antidamping depending on the optical, optomechanical, and thermal characteristics of the system.

\textit{Broadband multimode stabilization---}
The photothermal servo discussed above is predicted to enable broadband suppression of mechanical noise for all modes above a threshold frequency \cite{altin2017robust} while driving the optical cavity with a \textit{blue-detuned} optical spring. The environmentally driven noise variance is given by \cite{Miller2018Dec}
\begin{equation}\label{eq:variance}
\langle x^2 \rangle = \frac{1}{\Gamma_{\rm m}+ \Gamma_G+\Gamma_\mathcal{T}}\ \cdot \frac{ S_{\rm F}}{4 m^2 \Omega_{\rm m}^2},
\end{equation}
where $S_\text{F}$ is the power spectral density of the environmental force noise. 
As such, a change in the observed $\langle x^2 \rangle$ serves as a proxy for total optical damping ($\Gamma_G+\Gamma_\mathcal{T}$), as long as optical-spring induced changes in $\Omega_\text{m}$ and $m$ \cite{Clark2024Nov} are accounted for.

\begin{figure}[t!]
\centering
  \includegraphics[width=0.47\textwidth]{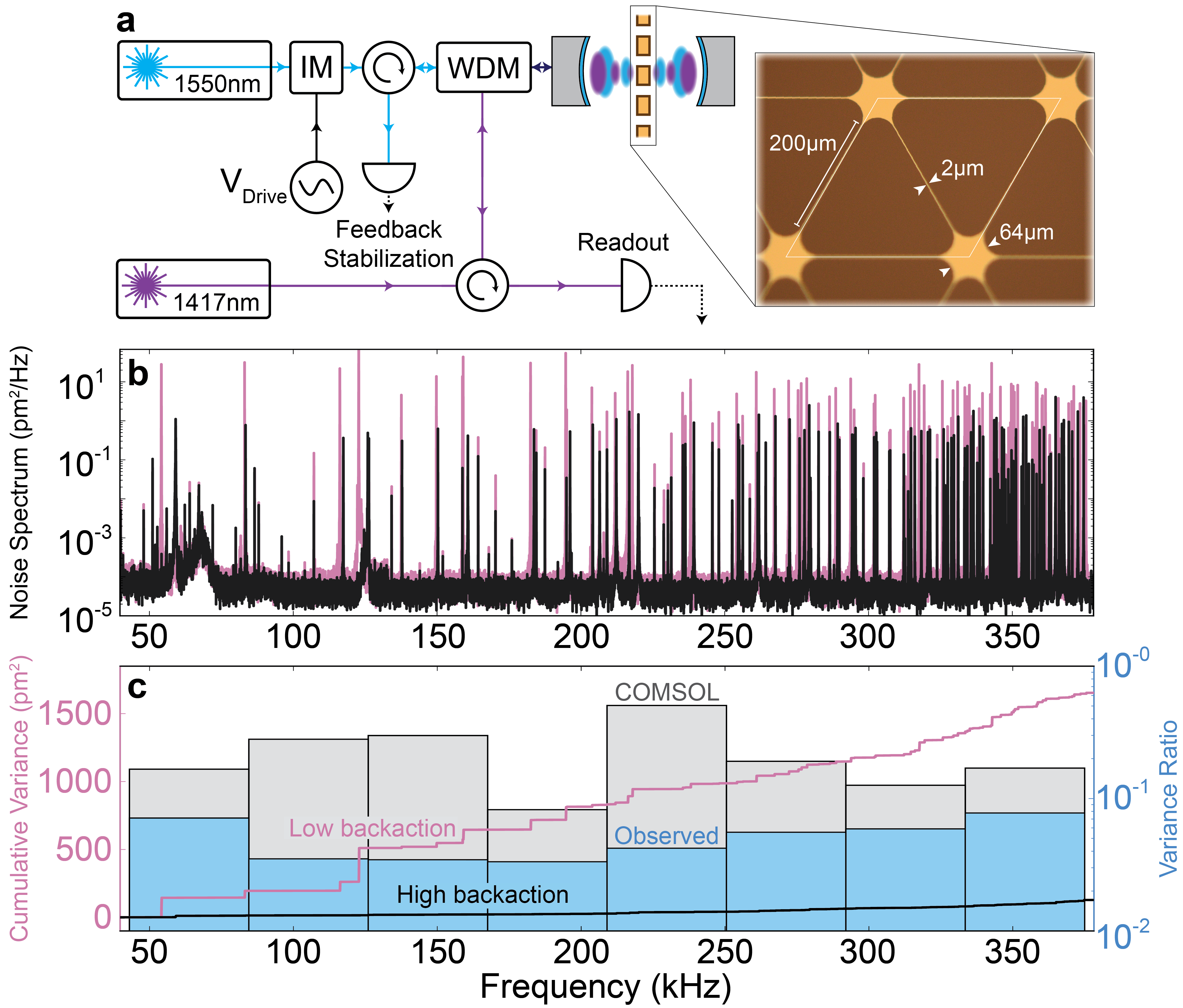}
  \caption{ \label{fig2} Blue-detuned multimode laser cooling.
  (a) Schematic. A 1550 nm ``control'' beam passes through an intensity modulator (IM) and circulator to a fiber cavity near the center of a patterned membrane (inset image). The control beam applies radiation pressure forces via the IM voltage $V_\text{Drive}$ and provides the error signal used to stabilize the cavity length via feedback to shear piezos attached to the mirrors. A 1417 nm beam provides a low-finesse, effectively backaction-free readout with fixed sensitivity, calibrated in (b) and (c) against the piezos' specified response to applied voltage; this provides an approximate reference scale but is not critical to our primary findings, as piezo response uncertainties introduce a common factor to both spectra, preserving their ratio. A wavelength division multiplexer (WDM) separates the two beams.
  (b) Displacement spectra recorded by the low-finesse 1417-nm beam. Pink shows the spectrum with the high-finesse beam (1550 nm, 0.5 mW input) blue-detuned to $\bar{\Delta}\sim 3\kappa$, where there is comparatively negligible back-action. Black shows the spectrum for $\bar{\Delta}=\kappa/\sqrt{12}$, where the optical spring is maximal and the membrane's mechanical noise peaks are suppressed. 
  (c) Pink and black curves show the cumulative variance of the spectra in (b). Blue bars are the ratio of cumulative variances for high and low backaction for smaller ranges of frequencies, while gray bars are simulated (COMSOL) variance ratio expected entirely from changes in mode mass, i.e., in the absence of optical damping. The comparatively low observed values (blue) provide a coarse summary of the optical damping's frequency dependence. 
  }
\end{figure}

Fig.~\ref{fig2}(a) shows the system used to observe multimode stabilization. The mechanical element is a 3.3~mm $\times$ 3.1~mm, 180 nm thick Si$_3$N$_4$ membrane (patterned into the shown hexagonal lattice) positioned within a 30-$\upmu$m-long fiber optical cavity addressed by two lasers. The first ``control'' laser (1550 nm) achieves finesse $10^4$ (depending on membrane position); the main purpose of this laser is to apply optomechanical forces, but its signal is also used to feedback-stabilize the cavity length via piezoelectric actuators below the fiber mirrors; note the feedback gain is reduced to a value well below the threshold at which it influences the mechanical spectrum (especially the damping parameters). The second ``readout'' laser (1417 nm) provides a low-finesse ($\lesssim 300$) readout with negligible back-action and sensitivity that is essentially independent of the control laser. 

The mirrors nominally comprise a 16-bilayer Bragg-stack -- each bilayer having a quarter-wavelength layer of SiO$_2$ (267 nm) and Ta$_2$O$_5$ (188 nm). Four bilayers are then removed from each \cite{Bernard2020Nov} to reduce the cavity finesse from their as-coated value $10^5$, leaving a nearly unetched Ta$_2$O$_5$ layer at the surface. Light circulating in the cavity applies radiation pressure over a $10$-$\upmu$m-diameter spot near the center of the membrane.
Figure~\ref{fig2}(b) shows two mechanical noise spectra recorded by the readout beam. 
The black spectrum is taken while the control beam is blue-detuned to $\bar{\Delta} = \kappa / \sqrt{12}$, i.e., where the optical spring is strongest and we expect large damping. The pink spectrum is recorded so far detuned ($\bar{\Delta }=3\kappa$) that the optical spring is reduced by a factor of $\sim$70 and the radiation-pressure (photothermal servo) damping by a factor $\gtrsim$2000 ($\gtrsim$5000). This approach allows us to closely approximate ``turning off'' the optomechanics without sacrificing the robust, consistent lock achieved with the (low-noise) control beam.
Importantly, in the high-backaction measurement, all prominent peaks  -- corresponding to the low mass, high quality factor membrane modes -- are suppressed: a first hallmark of broadband damping.

Figure~\ref{fig2}(c) shows the cumulative variance of the spectra in (b). The low-backaction curve (pink) exhibits pronounced steps at each membrane mode frequency. These are reduced at every step when the backaction is high (black), such that the variance from all modes (rightmost value) is suppressed by a factor of 13 over the full bandwidth. 
The blue bars show the ratio of integrated variances of the two spectra (including all modes) coarsely binned by frequency. In all bins, the remaining variance is more than a factor of 10 smaller. However, some of this suppression is known to arise from the increased stiffness and inertial mass of each mode as it is pinned under the optical spring \cite{Clark2024Nov}. To estimate this effect, the gray bins show a COMSOL simulation assuming $\Gamma_G=\Gamma_\mathcal{T}=0$ to provide an estimate of the suppressed variance expected \textit{exclusively} from mass and frequency changes. In all cases, the observed variance is reduced by more than this, indicating broadband damping. 

\begin{figure}[t!]
\centering
  \includegraphics[width=0.45\textwidth]{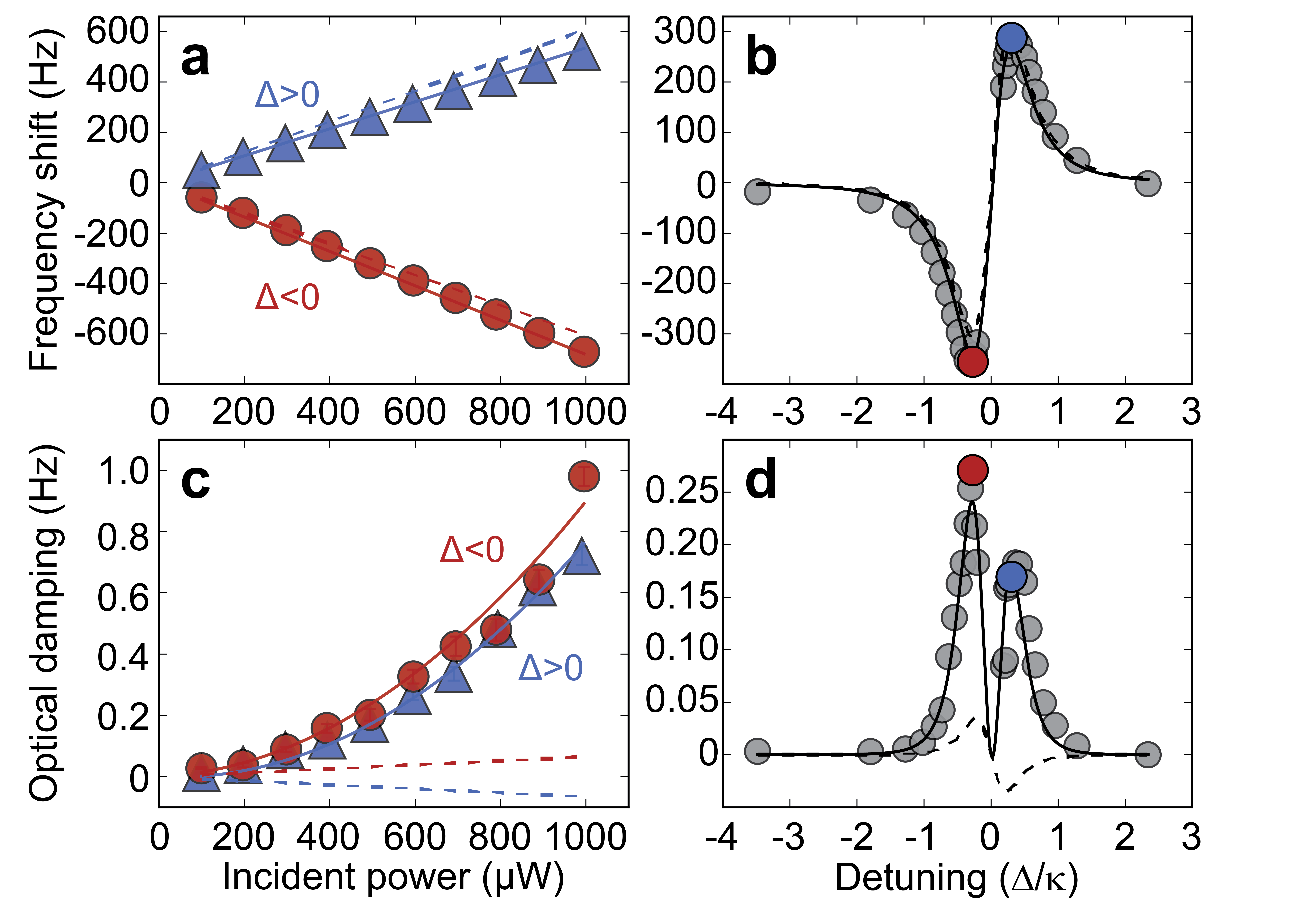}
  \caption{ \label{fig3} Power and detuning dependences of dynamical backaction including photothermal effects.
  (a) Frequency shift and (c) optical damping of the fundamental  (53.8 kHz) membrane mode versus incident power. For each power, the laser detuning is set to $\Delta \sim \pm \kappa / \sqrt{12}$, where the optical spring is maximal. Red (blue) dots correspond to laser detunings on the red (blue) side of resonance. Solid curves are simultaneous fits to the entire dataset including thermal loosening and radiation pressure contributions (see main text). 
  (b) Frequency shift and (d) optical damping of the fundamental mode as a function of laser detuning for an incident power $P_{\rm{in}}$=0.5 mW. 
  The dotted lines show the frequency shift and damping expected from radiation pressure alone using parameters extracted from the fit, highlighting the relative contribution of the photothermal servo. }
\end{figure}

\textit{Servo modified dynamics---}
To clearly distinguish between `force-like' dynamics \cite{Metzger2004Dec} (black and blue loops in Fig.~\ref{fig1}(b)) and `servo-like'  (red loop in Fig.~\ref{fig1}(b)),  we directly probe the changes in frequency (Fig.~\ref{fig3}(a,b)) and damping (Fig.~\ref{fig3}(c,d)) of the fundamental mechanical mode (53.8 kHz) as the control beam's power (at fixed detuning $\bar{\Delta}=\pm\kappa/\sqrt{12}$) and detuning (at fixed incident power 0.5 mW) are varied.
The (a) power and (b) detuning dependences of the frequency shift exhibit the linear power dependence and antisymmetric detuning dependence expected for either radiation-pressure dominated or servo-modified backaction, though with a gentle ``sag'' at higher $\bar{n}_\text{cav}$ due to the membrane heating, expanding, and loosening in proportion to intracavity power \cite{Clark2024Nov} (other modes follow similar behavior).  
However, the \textit{damping} in (c) exhibits a strong and positive \textit{quadratic} dependence on power, consistent with photothermal damping (Eq.~\ref{eq:Gamma_pt}), and  qualitatively different from the linear dependence (dashed lines) expected from radiation pressure alone. The detuning dependence in (d) also differs qualitatively from the radiation-pressure expectation, notably exhibiting positive damping for both signs of detuning.
The data in Fig.~\ref{fig3} is simultaneously fit (solid lines) to a model incorporating the sum of radiation pressure (Eqs.~\ref{eq:Kopt}-\ref{eq:Gamma_opt}), photothermal servo (Eq.~\ref{eq:Gamma_pt}), and loosening due to thermal expansion of the membrane material (modeled simply as a frequency shift $\delta\Omega_\text{th}=A/(\bar{\Delta}^2 / \kappa^2 +1/4 )$ in proportion to cavity power). Dashed lines show the contribution from radiation pressure alone for reference (i.e., $\text{Im}[\mathcal{T}]=0$). 
This fit reports optomechanical coupling strength $G = 1.4$ GHz/nm, thermal loosening coefficient $A=24$ Hz/mW incident, and $\text{Im}[\mathcal{T}]= 82$ Hz. The fit indicates optical losses of $(\kappa/2\pi, \kappa_{\rm in}/2\pi) =(3.0,0.2)$ GHz.
The agreement of this data with the servo-modified model -- along with the striking \textit{disagreement} with radiation-pressure-force expectations -- and the observed simultaneous noise suppression of the first $\gtrsim 100$ modes of a membrane represent the main results of this work.

\textit{Design considerations.---}
Whether thermal effects generate positive or negative damping depends on the competition between the thermoelastic expansion (transfer function $\mathcal{T}_\text{E}$), which shortens the cavity by growing the mirrors inward, and the thermorefractive index change (transfer function $\TR$), which effectively lengthens the cavity.
Combined with thermal lag at a given frequency $\omega$, the sum of these effects can generate damping only at frequencies $\omega$ where $\TR$ dominates \cite{Evans2008Nov, ballmer2015photothermal}. 
As $\omega$ decreases, the thermal penetration depth increases, leading to more material expanding and a larger value of $\TE$. By contrast, $\TR$ remains comparatively unchanged, since interference dictates that only the first few layers contain significant electric field. 
As such, there exists a ``crossover'' frequency below which thermal expansion dominates, resulting in antidamping. To realize the stability like what is shown here, it is critical to ensure this crossover is \textit{below} the fundamental frequency of the mechanical system. 
In previous free-space systems, the crossover frequency was simulated to be $\omega / 2\pi\sim$ 16 kHz for standard mirror coatings, resulting in antidamping of the (lowest-frequency) modes \cite{kelley2015observation}. 
In our system, the observed damping in Fig.~\ref{fig3} suggests that the crossover frequency is at most the fundamental mode frequency of 53.8 kHz, and $|\mathcal{T}|$ is large enough for the photothermal damping to dominate at modest laser powers.
One reason for this is likely our small cavity spot size ($\sim 10~\upmu$m diameter), which concentrates the heat deposition to an area $\gtrsim$~100 times smaller than that of a typical freespace cavity; this leads to faster thermal response, which can enhance the overall magnitude of $|\mathcal{T}|$ for a given frequency (as long as it's above the inverse thermal time scale \cite{cerdonio2001thermoelastic, deRosa2002experimental}).

System stability can in principle be further improved (and crossover frequency reduced) by increasing the relative contribution of $\TR$. Reference~\cite{kelley2015observation} proposes to achieve this using a very thick \SiO~layer on top of the coating, exploiting the fact that \SiO's thermo-refractive coefficient (8.5 ppm/K \cite{Leviton2006Jul}) is significantly higher than its thermo-elastic coefficient (0.51 ppm/K \cite{Fejer2004Oct}). Adding such thick layers to micron-scale fiber cavities is more cumbersome than for cm-scale mirrors -- diffraction precludes long fiber-fiber cavities -- but there are a few alternate approaches that can achieve the same goal. 
\begin{figure}[t]
\centering
\includegraphics[width=0.45\textwidth]{Fig4.png} 
  \caption{\label{fig4} Enhancement of the thermorefractive effect with modified mirror coatings.
  (a) Partial removal of the coating's terminating \TaO~layer enhances the electric field in the subsequent \SiO~layer, increasing $\TR$ by up to a factor of 2.3. Insets show the electric field profile (red) in the cavity (white) and first few layers of a Bragg mirror comprising \TaO~(blue) and \SiO~(gray) for 0, 53, 90\% removal. 
  (b) A half-wavelength \SiO~spacer embedded in the Bragg stack acts as a low-finesse ``mini-cavity", enhancing the field within it and $\TR$. Adding more layers after the spacer increases the mini-cavity finesse further boosting $\TR$, at the expense of added coating thickness. Note for simplicity we assume $\omega$ is low enough that the temperature profile is approximately uniform over the shown layers.}
\end{figure}
With existing \TaO-terminated mirror coatings, modest gains can be made by partially etching \cite{Bernard2020Nov} the topmost \TaO~layer. As shown in the inset of Fig.~\ref{fig4}(a), this serves to concentrate more light in the first \SiO~layer, thereby increasing $\TR$. While etching improves $\TR$ only by a factor of $\sim$2.3, this can lead to larger changes in the \textit{net} effect $\TR-\TE$, especially near the crossover frequency where $\TE$ and $\TR$ are balanced. This approach only slightly exceeds what is possible with a simple \SiO-terminated coating (corresponding to 100\% etched \TaO~in Fig.~\ref{fig4}(a)). However, it is advantageous with a freshly coated fiber mirror to etch the first few layers anyway, as this cleans the surface and removes the sidewall coating (so they can fit tight-tolerance ferrules, e.g.), and it practically easier to stop such an etch somewhere within the \TaO~layer \cite{Bernard2020Nov}. Note also that, as the \TaO~layer is removed, the electric field at the mirror surface increases, which can subsequently increase surface absorption.

Another promising approach is to design the mirror coating to enhance the field in the \SiO layer(s). Figure \ref{fig4}(b) shows an example of this, where the thickness of one glass ``spacer'' layer is doubled, creating a half-wavelength, low-finesse Fabry-Perot ``mini-cavity'' within the stack. This concentrates the field in the glass layer, enhancing $\TR$ as shown in the main plot. Embedding this ``mini-cavity'' deeper in the stack increases its finesse -- and hence the enhancement shown in Fig.~\ref{fig4}(b), so long as it is not deeper than thermal penetration depth. On the other hand, if heat is primarily deposited in the mini-cavity itself (where the field is high), then the thermal profile would follow, and very large enhancements are possible. Even in this case, however, the mini-cavity should not be embedded too deeply for two reasons. First, since the mini-cavity is (by design) resonant with the cavity light, it acts to transmit the incident cavity light through approximately twice the depth of the spacer layer (i.e., the mini-cavity comprises a comparable number of layers on \textit{both} sides of the spacer), thereby requiring additional dielectric layers to maintain the finesse of the main cavity. Second, a mini-cavity of effective length $l$ and finesse $\mathcal{F}$ will localize light outside the main cavity (see inset figures), thereby reducing the optical coupling to the object(s) of interest in the main cavity. Intuitively, whenever circulating photons pass through a mini-cavity, they travel an extra effective path length $\sim l\mathcal{F}$, thereby increasing the overall mode volume. This effect will be small so long as the main cavity length $L\gg l\mathcal{F}$, which should be straightforward for the shallow cavities shown in Fig.~\ref{fig4}(b).

\textit{Conclusions.---}
We demonstrate a cavity optical spring accompanied by broadband optical damping of the first $\gtrsim 100$ modes of a membrane in a fiber cavity. The broadband suppression of mechanical noise, along with the power-dependent and detuning-dependent damping and spring, is consistent with thermal dynamics in the mirror coatings that modify the cavity's susceptibility to detuning noise. %
This effect enables high-power measurement and stable optical spring without the antidamping-induced instabilities nominally associated with traditional radiation forces. Owing to its quadratic power dependence and reliance on thermal effects, this cooling technique is especially well-suited to higher-power applications and room temperature systems. 

\section{Acknowledgements}
We gratefully acknowledge initial fabrication support from Abeer Barasheed. TC acknowledges financial support from the Walter Sumner Fellowship. JCS acknowledges support from the Natural Sciences and Engineering Research Council of Canada (NSERC RGPIN 2018-05635), Canada Research Chairs (CRC 235060), Canadian foundation for Innovation (CFI 228130, 36423), Institut Transdisciplinaire d'Information Quantique (INTRIQ), and the Centre for the Physics of Materials (CPM) at McGill.

\section{Data Availability} 
The data presented here will be made available on the McGill Dataverse found at \\ https://borealisdata.ca/dataverse/mcgill.

% NOTE THIS LINE SHOULD BE REPLACED WITH 
 % ON SUBMISSION
%\bibliography{Bibliography}

\begin{thebibliography}{31}%
\makeatletter
\providecommand \@ifxundefined [1]{%
 \@ifx{#1\undefined}
}%
\providecommand \@ifnum [1]{%
 \ifnum #1\expandafter \@firstoftwo
 \else \expandafter \@secondoftwo
 \fi
}%
\providecommand \@ifx [1]{%
 \ifx #1\expandafter \@firstoftwo
 \else \expandafter \@secondoftwo
 \fi
}%
\providecommand \natexlab [1]{#1}%
\providecommand \enquote  [1]{``#1''}%
\providecommand \bibnamefont  [1]{#1}%
\providecommand \bibfnamefont [1]{#1}%
\providecommand \citenamefont [1]{#1}%
\providecommand \href@noop [0]{\@secondoftwo}%
\providecommand \href [0]{\begingroup \@sanitize@url \@href}%
\providecommand \@href[1]{\@@startlink{#1}\@@href}%
\providecommand \@@href[1]{\endgroup#1\@@endlink}%
\providecommand \@sanitize@url [0]{\catcode `\\12\catcode `\$12\catcode `\&12\catcode `\#12\catcode `\^12\catcode `\_12\catcode `\%12\relax}%
\providecommand \@@startlink[1]{}%
\providecommand \@@endlink[0]{}%
\providecommand \url  [0]{\begingroup\@sanitize@url \@url }%
\providecommand \@url [1]{\endgroup\@href {#1}{\urlprefix }}%
\providecommand \urlprefix  [0]{URL }%
\providecommand \Eprint [0]{\href }%
\providecommand \doibase [0]{https://doi.org/}%
\providecommand \selectlanguage [0]{\@gobble}%
\providecommand \bibinfo  [0]{\@secondoftwo}%
\providecommand \bibfield  [0]{\@secondoftwo}%
\providecommand \translation [1]{[#1]}%
\providecommand \BibitemOpen [0]{}%
\providecommand \bibitemStop [0]{}%
\providecommand \bibitemNoStop [0]{.\EOS\space}%
\providecommand \EOS [0]{\spacefactor3000\relax}%
\providecommand \BibitemShut  [1]{\csname bibitem#1\endcsname}%
\let\auto@bib@innerbib\@empty
%</preamble>
\bibitem [{\citenamefont {Aspelmeyer}\ \emph {et~al.}(2014)\citenamefont {Aspelmeyer}, \citenamefont {Kippenberg},\ and\ \citenamefont {Marquardt}}]{Aspelmeyer2014Dec}%
  \BibitemOpen
  \bibfield  {author} {\bibinfo {author} {\bibfnamefont {M.}~\bibnamefont {Aspelmeyer}}, \bibinfo {author} {\bibfnamefont {T.~J.}\ \bibnamefont {Kippenberg}},\ and\ \bibinfo {author} {\bibfnamefont {F.}~\bibnamefont {Marquardt}},\ }\bibfield  {title} {\bibinfo {title} {{Cavity optomechanics}},\ }\href {https://doi.org/10.1103/RevModPhys.86.1391} {\bibfield  {journal} {\bibinfo  {journal} {Rev. Mod. Phys.}\ }\textbf {\bibinfo {volume} {86}},\ \bibinfo {pages} {1391} (\bibinfo {year} {2014})}\BibitemShut {NoStop}%
\bibitem [{\citenamefont {Aasi}\ \emph {et~al.}(2013)\citenamefont {Aasi}, \citenamefont {Abadie}, \citenamefont {Abbott},\ and\ \citenamefont {et~al.}}]{Aasi2013Aug}%
  \BibitemOpen
  \bibfield  {author} {\bibinfo {author} {\bibfnamefont {J.}~\bibnamefont {Aasi}}, \bibinfo {author} {\bibfnamefont {J.}~\bibnamefont {Abadie}}, \bibinfo {author} {\bibfnamefont {B.~P.}\ \bibnamefont {Abbott}},\ and\ \bibinfo {author} {\bibnamefont {et~al.}},\ }\bibfield  {title} {\bibinfo {title} {Enhanced sensitivity of the ligo gravitational wave detector by using squeezed states of light},\ }\href {https://doi.org/10.1038/nphoton.2013.177} {\bibfield  {journal} {\bibinfo  {journal} {Nat. Photonics}\ }\textbf {\bibinfo {volume} {7}},\ \bibinfo {pages} {613} (\bibinfo {year} {2013})}\BibitemShut {NoStop}%
\bibitem [{\citenamefont {Cripe}\ \emph {et~al.}(2019)\citenamefont {Cripe}, \citenamefont {Aggarwal}, \citenamefont {Lanza}, \citenamefont {Libson}, \citenamefont {Singh}, \citenamefont {Heu}, \citenamefont {Follman}, \citenamefont {Cole}, \citenamefont {Mavalvala},\ and\ \citenamefont {Corbitt}}]{Cripe2019Apr}%
  \BibitemOpen
  \bibfield  {author} {\bibinfo {author} {\bibfnamefont {J.}~\bibnamefont {Cripe}}, \bibinfo {author} {\bibfnamefont {N.}~\bibnamefont {Aggarwal}}, \bibinfo {author} {\bibfnamefont {R.}~\bibnamefont {Lanza}}, \bibinfo {author} {\bibfnamefont {A.}~\bibnamefont {Libson}}, \bibinfo {author} {\bibfnamefont {R.}~\bibnamefont {Singh}}, \bibinfo {author} {\bibfnamefont {P.}~\bibnamefont {Heu}}, \bibinfo {author} {\bibfnamefont {D.}~\bibnamefont {Follman}}, \bibinfo {author} {\bibfnamefont {G.~D.}\ \bibnamefont {Cole}}, \bibinfo {author} {\bibfnamefont {N.}~\bibnamefont {Mavalvala}},\ and\ \bibinfo {author} {\bibfnamefont {T.}~\bibnamefont {Corbitt}},\ }\bibfield  {title} {\bibinfo {title} {{Measurement of quantum back action in the audio band at room temperature}},\ }\href {https://doi.org/10.1038/s41586-019-1051-4} {\bibfield  {journal} {\bibinfo  {journal} {Nature}\ }\textbf {\bibinfo {volume} {568}},\ \bibinfo {pages} {364} (\bibinfo {year} {2019})}\BibitemShut {NoStop}%
\bibitem [{\citenamefont {Aggarwal}\ \emph {et~al.}(2020)\citenamefont {Aggarwal}, \citenamefont {Cullen}, \citenamefont {Cripe}, \citenamefont {Cole}, \citenamefont {Lanza}, \citenamefont {Libson}, \citenamefont {Follman}, \citenamefont {Heu}, \citenamefont {Corbitt},\ and\ \citenamefont {Mavalvala}}]{Aggarwal2020Jul}%
  \BibitemOpen
  \bibfield  {author} {\bibinfo {author} {\bibfnamefont {N.}~\bibnamefont {Aggarwal}}, \bibinfo {author} {\bibfnamefont {T.~J.}\ \bibnamefont {Cullen}}, \bibinfo {author} {\bibfnamefont {J.}~\bibnamefont {Cripe}}, \bibinfo {author} {\bibfnamefont {G.~D.}\ \bibnamefont {Cole}}, \bibinfo {author} {\bibfnamefont {R.}~\bibnamefont {Lanza}}, \bibinfo {author} {\bibfnamefont {A.}~\bibnamefont {Libson}}, \bibinfo {author} {\bibfnamefont {D.}~\bibnamefont {Follman}}, \bibinfo {author} {\bibfnamefont {P.}~\bibnamefont {Heu}}, \bibinfo {author} {\bibfnamefont {T.}~\bibnamefont {Corbitt}},\ and\ \bibinfo {author} {\bibfnamefont {N.}~\bibnamefont {Mavalvala}},\ }\bibfield  {title} {\bibinfo {title} {{Room-temperature optomechanical squeezing}},\ }\href {https://doi.org/10.1038/s41567-020-0877-x} {\bibfield  {journal} {\bibinfo  {journal} {Nat. Phys.}\ }\textbf {\bibinfo {volume} {16}},\ \bibinfo {pages} {784} (\bibinfo {year} {2020})}\BibitemShut {NoStop}%
\bibitem [{\citenamefont {Saulson}(1990)}]{saulson1990thermal}%
  \BibitemOpen
  \bibfield  {author} {\bibinfo {author} {\bibfnamefont {P.~R.}\ \bibnamefont {Saulson}},\ }\bibfield  {title} {\bibinfo {title} {Thermal noise in mechanical experiments},\ }\href@noop {} {\bibfield  {journal} {\bibinfo  {journal} {Physical Review D}\ }\textbf {\bibinfo {volume} {42}},\ \bibinfo {pages} {2437} (\bibinfo {year} {1990})}\BibitemShut {NoStop}%
\bibitem [{\citenamefont {Fedorov}\ \emph {et~al.}(2020)\citenamefont {Fedorov}, \citenamefont {Fedorov}, \citenamefont {Beccari}, \citenamefont {Beccari}, \citenamefont {Arabmoheghi}, \citenamefont {Wilson}, \citenamefont {Engelsen}, \citenamefont {Engelsen},\ and\ \citenamefont {Kippenberg}}]{Fedorov2020Nov}%
  \BibitemOpen
  \bibfield  {author} {\bibinfo {author} {\bibfnamefont {S.~A.}\ \bibnamefont {Fedorov}}, \bibinfo {author} {\bibfnamefont {S.~A.}\ \bibnamefont {Fedorov}}, \bibinfo {author} {\bibfnamefont {A.}~\bibnamefont {Beccari}}, \bibinfo {author} {\bibfnamefont {A.}~\bibnamefont {Beccari}}, \bibinfo {author} {\bibfnamefont {A.}~\bibnamefont {Arabmoheghi}}, \bibinfo {author} {\bibfnamefont {D.~J.}\ \bibnamefont {Wilson}}, \bibinfo {author} {\bibfnamefont {N.~J.}\ \bibnamefont {Engelsen}}, \bibinfo {author} {\bibfnamefont {N.~J.}\ \bibnamefont {Engelsen}},\ and\ \bibinfo {author} {\bibfnamefont {T.~J.}\ \bibnamefont {Kippenberg}},\ }\bibfield  {title} {\bibinfo {title} {{Thermal intermodulation noise in cavity-based measurements}},\ }\href {https://doi.org/10.1364/OPTICA.402449} {\bibfield  {journal} {\bibinfo  {journal} {Optica}\ }\textbf {\bibinfo {volume} {7}},\ \bibinfo {pages} {1609} (\bibinfo {year} {2020})}\BibitemShut {NoStop}%
\bibitem [{\citenamefont {Pluchar}\ \emph {et~al.}(2023)\citenamefont {Pluchar}, \citenamefont {Agrawal},\ and\ \citenamefont {Wilson}}]{Pluchar2023Nov}%
  \BibitemOpen
  \bibfield  {author} {\bibinfo {author} {\bibfnamefont {C.~M.}\ \bibnamefont {Pluchar}}, \bibinfo {author} {\bibfnamefont {A.~R.}\ \bibnamefont {Agrawal}},\ and\ \bibinfo {author} {\bibfnamefont {D.~J.}\ \bibnamefont {Wilson}},\ }\bibfield  {title} {\bibinfo {title} {{Thermal intermodulation backaction in a high-cooperativity optomechanical system}},\ }\href {https://doi.org/10.1364/OPTICA.500123} {\bibfield  {journal} {\bibinfo  {journal} {Optica}\ }\textbf {\bibinfo {volume} {10}},\ \bibinfo {pages} {1543} (\bibinfo {year} {2023})}\BibitemShut {NoStop}%
\bibitem [{\citenamefont {Tsaturyan}\ \emph {et~al.}(2017)\citenamefont {Tsaturyan}, \citenamefont {Barg}, \citenamefont {Polzik},\ and\ \citenamefont {Schliesser}}]{Tsaturyan2017Aug}%
  \BibitemOpen
  \bibfield  {author} {\bibinfo {author} {\bibfnamefont {Y.}~\bibnamefont {Tsaturyan}}, \bibinfo {author} {\bibfnamefont {A.}~\bibnamefont {Barg}}, \bibinfo {author} {\bibfnamefont {E.~S.}\ \bibnamefont {Polzik}},\ and\ \bibinfo {author} {\bibfnamefont {A.}~\bibnamefont {Schliesser}},\ }\bibfield  {title} {\bibinfo {title} {{Ultracoherent nanomechanical resonators via soft clamping and dissipation dilution}},\ }\href {https://doi.org/10.1038/nnano.2017.101} {\bibfield  {journal} {\bibinfo  {journal} {Nat. Nanotechnol.}\ }\textbf {\bibinfo {volume} {12}},\ \bibinfo {pages} {776} (\bibinfo {year} {2017})}\BibitemShut {NoStop}%
\bibitem [{\citenamefont {Mason}\ \emph {et~al.}(2019)\citenamefont {Mason}, \citenamefont {Chen}, \citenamefont {Rossi}, \citenamefont {Tsaturyan},\ and\ \citenamefont {Schliesser}}]{Mason2019Aug}%
  \BibitemOpen
  \bibfield  {author} {\bibinfo {author} {\bibfnamefont {D.}~\bibnamefont {Mason}}, \bibinfo {author} {\bibfnamefont {J.}~\bibnamefont {Chen}}, \bibinfo {author} {\bibfnamefont {M.}~\bibnamefont {Rossi}}, \bibinfo {author} {\bibfnamefont {Y.}~\bibnamefont {Tsaturyan}},\ and\ \bibinfo {author} {\bibfnamefont {A.}~\bibnamefont {Schliesser}},\ }\bibfield  {title} {\bibinfo {title} {{Continuous force and displacement measurement below the standard quantum limit}},\ }\href {https://doi.org/10.1038/s41567-019-0533-5} {\bibfield  {journal} {\bibinfo  {journal} {Nat. Phys.}\ }\textbf {\bibinfo {volume} {15}},\ \bibinfo {pages} {745} (\bibinfo {year} {2019})}\BibitemShut {NoStop}%
\bibitem [{\citenamefont {Huang}\ \emph {et~al.}(2024)\citenamefont {Huang}, \citenamefont {Beccari}, \citenamefont {Engelsen},\ and\ \citenamefont {Kippenberg}}]{Huang2024Feb}%
  \BibitemOpen
  \bibfield  {author} {\bibinfo {author} {\bibfnamefont {G.}~\bibnamefont {Huang}}, \bibinfo {author} {\bibfnamefont {A.}~\bibnamefont {Beccari}}, \bibinfo {author} {\bibfnamefont {N.~J.}\ \bibnamefont {Engelsen}},\ and\ \bibinfo {author} {\bibfnamefont {T.~J.}\ \bibnamefont {Kippenberg}},\ }\bibfield  {title} {\bibinfo {title} {{Room-temperature quantum optomechanics using an ultralow noise cavity}},\ }\href {https://doi.org/10.1038/s41586-023-06997-3} {\bibfield  {journal} {\bibinfo  {journal} {Nature}\ }\textbf {\bibinfo {volume} {626}},\ \bibinfo {pages} {512} (\bibinfo {year} {2024})}\BibitemShut {NoStop}%
\bibitem [{\citenamefont {Rossi}\ \emph {et~al.}(2018)\citenamefont {Rossi}, \citenamefont {Mason}, \citenamefont {Chen}, \citenamefont {Tsaturyan},\ and\ \citenamefont {Schliesser}}]{Rossi2018Nov}%
  \BibitemOpen
  \bibfield  {author} {\bibinfo {author} {\bibfnamefont {M.}~\bibnamefont {Rossi}}, \bibinfo {author} {\bibfnamefont {D.}~\bibnamefont {Mason}}, \bibinfo {author} {\bibfnamefont {J.}~\bibnamefont {Chen}}, \bibinfo {author} {\bibfnamefont {Y.}~\bibnamefont {Tsaturyan}},\ and\ \bibinfo {author} {\bibfnamefont {A.}~\bibnamefont {Schliesser}},\ }\bibfield  {title} {\bibinfo {title} {{Measurement-based quantum control of mechanical motion}},\ }\href {https://doi.org/10.1038/s41586-018-0643-8} {\bibfield  {journal} {\bibinfo  {journal} {Nature}\ }\textbf {\bibinfo {volume} {563}},\ \bibinfo {pages} {53} (\bibinfo {year} {2018})}\BibitemShut {NoStop}%
\bibitem [{\citenamefont {Cuthbertson}\ \emph {et~al.}(1996)\citenamefont {Cuthbertson}, \citenamefont {Tobar}, \citenamefont {Ivanov},\ and\ \citenamefont {Blair}}]{Cuthbertson1996Jul}%
  \BibitemOpen
  \bibfield  {author} {\bibinfo {author} {\bibfnamefont {B.~D.}\ \bibnamefont {Cuthbertson}}, \bibinfo {author} {\bibfnamefont {M.~E.}\ \bibnamefont {Tobar}}, \bibinfo {author} {\bibfnamefont {E.~N.}\ \bibnamefont {Ivanov}},\ and\ \bibinfo {author} {\bibfnamefont {D.~G.}\ \bibnamefont {Blair}},\ }\bibfield  {title} {\bibinfo {title} {{Parametric back{-}action effects in a high{-}Q cyrogenic sapphire transducer}},\ }\href {https://doi.org/10.1063/1.1147193} {\bibfield  {journal} {\bibinfo  {journal} {Rev. Sci. Instrum.}\ }\textbf {\bibinfo {volume} {67}},\ \bibinfo {pages} {2435} (\bibinfo {year} {1996})}\BibitemShut {NoStop}%
\bibitem [{\citenamefont {Sheard}\ \emph {et~al.}(2004)\citenamefont {Sheard}, \citenamefont {Gray}, \citenamefont {Mow-Lowry}, \citenamefont {McClelland},\ and\ \citenamefont {Whitcomb}}]{Sheard2004May}%
  \BibitemOpen
  \bibfield  {author} {\bibinfo {author} {\bibfnamefont {B.~S.}\ \bibnamefont {Sheard}}, \bibinfo {author} {\bibfnamefont {M.~B.}\ \bibnamefont {Gray}}, \bibinfo {author} {\bibfnamefont {C.~M.}\ \bibnamefont {Mow-Lowry}}, \bibinfo {author} {\bibfnamefont {D.~E.}\ \bibnamefont {McClelland}},\ and\ \bibinfo {author} {\bibfnamefont {S.~E.}\ \bibnamefont {Whitcomb}},\ }\bibfield  {title} {\bibinfo {title} {{Observation and characterization of an optical spring}},\ }\href {https://doi.org/10.1103/PhysRevA.69.051801} {\bibfield  {journal} {\bibinfo  {journal} {Phys. Rev. A}\ }\textbf {\bibinfo {volume} {69}},\ \bibinfo {pages} {051801} (\bibinfo {year} {2004})}\BibitemShut {NoStop}%
\bibitem [{\citenamefont {Arcizet}\ \emph {et~al.}(2006)\citenamefont {Arcizet}, \citenamefont {Cohadon}, \citenamefont {Briant}, \citenamefont {Pinard},\ and\ \citenamefont {Heidmann}}]{Arcizet2006Nov}%
  \BibitemOpen
  \bibfield  {author} {\bibinfo {author} {\bibfnamefont {O.}~\bibnamefont {Arcizet}}, \bibinfo {author} {\bibfnamefont {P.-F.}\ \bibnamefont {Cohadon}}, \bibinfo {author} {\bibfnamefont {T.}~\bibnamefont {Briant}}, \bibinfo {author} {\bibfnamefont {M.}~\bibnamefont {Pinard}},\ and\ \bibinfo {author} {\bibfnamefont {A.}~\bibnamefont {Heidmann}},\ }\bibfield  {title} {\bibinfo {title} {{Radiation-pressure cooling and optomechanical instability of a micromirror}},\ }\href {https://doi.org/10.1038/nature05244} {\bibfield  {journal} {\bibinfo  {journal} {Nature}\ }\textbf {\bibinfo {volume} {444}},\ \bibinfo {pages} {71} (\bibinfo {year} {2006})}\BibitemShut {NoStop}%
\bibitem [{\citenamefont {Corbitt}\ \emph {et~al.}(2007{\natexlab{a}})\citenamefont {Corbitt}, \citenamefont {Chen}, \citenamefont {Innerhofer}, \citenamefont {M{\ifmmode\ddot{u}\else\"{u}\fi}ller-Ebhardt}, \citenamefont {Ottaway}, \citenamefont {Rehbein}, \citenamefont {Sigg}, \citenamefont {Whitcomb}, \citenamefont {Wipf},\ and\ \citenamefont {Mavalvala}}]{corbitt2007all}%
  \BibitemOpen
  \bibfield  {author} {\bibinfo {author} {\bibfnamefont {T.}~\bibnamefont {Corbitt}}, \bibinfo {author} {\bibfnamefont {Y.}~\bibnamefont {Chen}}, \bibinfo {author} {\bibfnamefont {E.}~\bibnamefont {Innerhofer}}, \bibinfo {author} {\bibfnamefont {H.}~\bibnamefont {M{\ifmmode\ddot{u}\else\"{u}\fi}ller-Ebhardt}}, \bibinfo {author} {\bibfnamefont {D.}~\bibnamefont {Ottaway}}, \bibinfo {author} {\bibfnamefont {H.}~\bibnamefont {Rehbein}}, \bibinfo {author} {\bibfnamefont {D.}~\bibnamefont {Sigg}}, \bibinfo {author} {\bibfnamefont {S.}~\bibnamefont {Whitcomb}}, \bibinfo {author} {\bibfnamefont {C.}~\bibnamefont {Wipf}},\ and\ \bibinfo {author} {\bibfnamefont {N.}~\bibnamefont {Mavalvala}},\ }\bibfield  {title} {\bibinfo {title} {{An All-Optical Trap for a Gram-Scale Mirror}},\ }\href {https://doi.org/10.1103/PhysRevLett.98.150802} {\bibfield  {journal} {\bibinfo  {journal} {Phys. Rev. Lett.}\ }\textbf {\bibinfo {volume} {98}},\ \bibinfo {pages} {150802} (\bibinfo {year} {2007}{\natexlab{a}})}\BibitemShut {NoStop}%
\bibitem [{\citenamefont {Singh}\ \emph {et~al.}(2016)\citenamefont {Singh}, \citenamefont {Cole}, \citenamefont {Cripe},\ and\ \citenamefont {Corbitt}}]{Singh2016Nov}%
  \BibitemOpen
  \bibfield  {author} {\bibinfo {author} {\bibfnamefont {R.}~\bibnamefont {Singh}}, \bibinfo {author} {\bibfnamefont {G.~D.}\ \bibnamefont {Cole}}, \bibinfo {author} {\bibfnamefont {J.}~\bibnamefont {Cripe}},\ and\ \bibinfo {author} {\bibfnamefont {T.}~\bibnamefont {Corbitt}},\ }\bibfield  {title} {\bibinfo {title} {{Stable Optical Trap from a Single Optical Field Utilizing Birefringence}},\ }\href {https://doi.org/10.1103/PhysRevLett.117.213604} {\bibfield  {journal} {\bibinfo  {journal} {Phys. Rev. Lett.}\ }\textbf {\bibinfo {volume} {117}},\ \bibinfo {pages} {213604} (\bibinfo {year} {2016})}\BibitemShut {NoStop}%
\bibitem [{\citenamefont {Corbitt}\ \emph {et~al.}(2007{\natexlab{b}})\citenamefont {Corbitt}, \citenamefont {Wipf}, \citenamefont {Bodiya}, \citenamefont {Ottaway}, \citenamefont {Sigg}, \citenamefont {Smith}, \citenamefont {Whitcomb},\ and\ \citenamefont {Mavalvala}}]{Corbitt2007Oct}%
  \BibitemOpen
  \bibfield  {author} {\bibinfo {author} {\bibfnamefont {T.}~\bibnamefont {Corbitt}}, \bibinfo {author} {\bibfnamefont {C.}~\bibnamefont {Wipf}}, \bibinfo {author} {\bibfnamefont {T.}~\bibnamefont {Bodiya}}, \bibinfo {author} {\bibfnamefont {D.}~\bibnamefont {Ottaway}}, \bibinfo {author} {\bibfnamefont {D.}~\bibnamefont {Sigg}}, \bibinfo {author} {\bibfnamefont {N.}~\bibnamefont {Smith}}, \bibinfo {author} {\bibfnamefont {S.}~\bibnamefont {Whitcomb}},\ and\ \bibinfo {author} {\bibfnamefont {N.}~\bibnamefont {Mavalvala}},\ }\bibfield  {title} {\bibinfo {title} {{Optical Dilution and Feedback Cooling of a Gram-Scale Oscillator to 6.9 mK}},\ }\href {https://doi.org/10.1103/PhysRevLett.99.160801} {\bibfield  {journal} {\bibinfo  {journal} {Phys. Rev. Lett.}\ }\textbf {\bibinfo {volume} {99}},\ \bibinfo {pages} {160801} (\bibinfo {year} {2007}{\natexlab{b}})}\BibitemShut {NoStop}%
\bibitem [{\citenamefont {Mow-Lowry}\ \emph {et~al.}(2008)\citenamefont {Mow-Lowry}, \citenamefont {Mullavey}, \citenamefont {Go{\ss}ler}, \citenamefont {Gray},\ and\ \citenamefont {McClelland}}]{mow2008cooling}%
  \BibitemOpen
  \bibfield  {author} {\bibinfo {author} {\bibfnamefont {C.}~\bibnamefont {Mow-Lowry}}, \bibinfo {author} {\bibfnamefont {A.}~\bibnamefont {Mullavey}}, \bibinfo {author} {\bibfnamefont {S.}~\bibnamefont {Go{\ss}ler}}, \bibinfo {author} {\bibfnamefont {M.~B.}\ \bibnamefont {Gray}},\ and\ \bibinfo {author} {\bibfnamefont {D.}~\bibnamefont {McClelland}},\ }\bibfield  {title} {\bibinfo {title} {Cooling of a gram-scale cantilever flexure to 70 mk with a servo-modified optical spring},\ }\href {https://doi.org/10.1103/PhysRevLett.100.010801} {\bibfield  {journal} {\bibinfo  {journal} {Phys. Rev. Lett.}\ }\textbf {\bibinfo {volume} {100}},\ \bibinfo {pages} {010801} (\bibinfo {year} {2008})}\BibitemShut {NoStop}%
\bibitem [{\citenamefont {Schediwy}\ \emph {et~al.}(2008)\citenamefont {Schediwy}, \citenamefont {Zhao}, \citenamefont {Ju}, \citenamefont {Blair},\ and\ \citenamefont {Willems}}]{Schediwy2008Jan}%
  \BibitemOpen
  \bibfield  {author} {\bibinfo {author} {\bibfnamefont {S.~W.}\ \bibnamefont {Schediwy}}, \bibinfo {author} {\bibfnamefont {C.}~\bibnamefont {Zhao}}, \bibinfo {author} {\bibfnamefont {L.}~\bibnamefont {Ju}}, \bibinfo {author} {\bibfnamefont {D.~G.}\ \bibnamefont {Blair}},\ and\ \bibinfo {author} {\bibfnamefont {P.}~\bibnamefont {Willems}},\ }\bibfield  {title} {\bibinfo {title} {{Observation of enhanced optical spring damping in a macroscopic mechanical resonator and application for parametric instability control in advanced gravitational-wave detectors}},\ }\href {https://doi.org/10.1103/PhysRevA.77.013813} {\bibfield  {journal} {\bibinfo  {journal} {Phys. Rev. A}\ }\textbf {\bibinfo {volume} {77}},\ \bibinfo {pages} {013813} (\bibinfo {year} {2008})}\BibitemShut {NoStop}%
\bibitem [{\citenamefont {Ballmer}(2015)}]{ballmer2015photothermal}%
  \BibitemOpen
  \bibfield  {author} {\bibinfo {author} {\bibfnamefont {S.~W.}\ \bibnamefont {Ballmer}},\ }\bibfield  {title} {\bibinfo {title} {{Photothermal transfer function of dielectric mirrors for precision measurements}},\ }\href {https://doi.org/10.1103/PhysRevD.91.023010} {\bibfield  {journal} {\bibinfo  {journal} {Phys. Rev. D}\ }\textbf {\bibinfo {volume} {91}},\ \bibinfo {pages} {023010} (\bibinfo {year} {2015})}\BibitemShut {NoStop}%
\bibitem [{\citenamefont {Kelley}\ \emph {et~al.}(2015)\citenamefont {Kelley}, \citenamefont {Lough}, \citenamefont {Manga{\ifmmode\tilde{n}\else\~{n}\fi}a-Sandoval}, \citenamefont {Perreca},\ and\ \citenamefont {Ballmer}}]{kelley2015observation}%
  \BibitemOpen
  \bibfield  {author} {\bibinfo {author} {\bibfnamefont {D.}~\bibnamefont {Kelley}}, \bibinfo {author} {\bibfnamefont {J.}~\bibnamefont {Lough}}, \bibinfo {author} {\bibfnamefont {F.}~\bibnamefont {Manga{\ifmmode\tilde{n}\else\~{n}\fi}a-Sandoval}}, \bibinfo {author} {\bibfnamefont {A.}~\bibnamefont {Perreca}},\ and\ \bibinfo {author} {\bibfnamefont {S.~W.}\ \bibnamefont {Ballmer}},\ }\bibfield  {title} {\bibinfo {title} {{Observation of photothermal feedback in a stable dual-carrier optical spring}},\ }\href {https://doi.org/10.1103/PhysRevD.92.062003} {\bibfield  {journal} {\bibinfo  {journal} {Phys. Rev. D}\ }\textbf {\bibinfo {volume} {92}},\ \bibinfo {pages} {062003} (\bibinfo {year} {2015})}\BibitemShut {NoStop}%
\bibitem [{\citenamefont {Altin}\ \emph {et~al.}(2017)\citenamefont {Altin}, \citenamefont {Nguyen}, \citenamefont {Slagmolen}, \citenamefont {Ward}, \citenamefont {Shaddock},\ and\ \citenamefont {McClelland}}]{altin2017robust}%
  \BibitemOpen
  \bibfield  {author} {\bibinfo {author} {\bibfnamefont {P.~A.}\ \bibnamefont {Altin}}, \bibinfo {author} {\bibfnamefont {T.~T.-H.}\ \bibnamefont {Nguyen}}, \bibinfo {author} {\bibfnamefont {B.~J.~J.}\ \bibnamefont {Slagmolen}}, \bibinfo {author} {\bibfnamefont {R.~L.}\ \bibnamefont {Ward}}, \bibinfo {author} {\bibfnamefont {D.~A.}\ \bibnamefont {Shaddock}},\ and\ \bibinfo {author} {\bibfnamefont {D.~E.}\ \bibnamefont {McClelland}},\ }\bibfield  {title} {\bibinfo {title} {{A robust single-beam optical trap for a gram-scale mechanical oscillator}},\ }\href {https://doi.org/10.1038/s41598-017-15179-x} {\bibfield  {journal} {\bibinfo  {journal} {Sci. Rep.}\ }\textbf {\bibinfo {volume} {7}},\ \bibinfo {pages} {1} (\bibinfo {year} {2017})}\BibitemShut {NoStop}%
\bibitem [{\citenamefont {Bernard}\ \emph {et~al.}(2020)\citenamefont {Bernard}, \citenamefont {Bernard}, \citenamefont {Clark}, \citenamefont {Clark}, \citenamefont {Dumont}, \citenamefont {Ma},\ and\ \citenamefont {Sankey}}]{Bernard2020Nov}%
  \BibitemOpen
  \bibfield  {author} {\bibinfo {author} {\bibfnamefont {S.}~\bibnamefont {Bernard}}, \bibinfo {author} {\bibfnamefont {S.}~\bibnamefont {Bernard}}, \bibinfo {author} {\bibfnamefont {T.~J.}\ \bibnamefont {Clark}}, \bibinfo {author} {\bibfnamefont {T.~J.}\ \bibnamefont {Clark}}, \bibinfo {author} {\bibfnamefont {V.}~\bibnamefont {Dumont}}, \bibinfo {author} {\bibfnamefont {J.}~\bibnamefont {Ma}},\ and\ \bibinfo {author} {\bibfnamefont {J.~C.}\ \bibnamefont {Sankey}},\ }\bibfield  {title} {\bibinfo {title} {{Monitored wet-etch removal of individual dielectric layers from high-finesse Bragg mirrors}},\ }\href {https://doi.org/10.1364/OE.400986} {\bibfield  {journal} {\bibinfo  {journal} {Opt. Express}\ }\textbf {\bibinfo {volume} {28}},\ \bibinfo {pages} {33823} (\bibinfo {year} {2020})}\BibitemShut {NoStop}%
\bibitem [{\citenamefont {Metzger}\ and\ \citenamefont {Karrai}(2004)}]{Metzger2004Dec}%
  \BibitemOpen
  \bibfield  {author} {\bibinfo {author} {\bibfnamefont {C.~H.}\ \bibnamefont {Metzger}}\ and\ \bibinfo {author} {\bibfnamefont {K.}~\bibnamefont {Karrai}},\ }\bibfield  {title} {\bibinfo {title} {{Cavity cooling of a microlever}},\ }\href {https://doi.org/10.1038/nature03118} {\bibfield  {journal} {\bibinfo  {journal} {Nature}\ }\textbf {\bibinfo {volume} {432}},\ \bibinfo {pages} {1002} (\bibinfo {year} {2004})}\BibitemShut {NoStop}%
\bibitem [{\citenamefont {Miller}\ \emph {et~al.}(2018)\citenamefont {Miller}, \citenamefont {Ansari}, \citenamefont {Heinz}, \citenamefont {Chen}, \citenamefont {Flader}, \citenamefont {Shin}, \citenamefont {Villanueva},\ and\ \citenamefont {Kenny}}]{Miller2018Dec}%
  \BibitemOpen
  \bibfield  {author} {\bibinfo {author} {\bibfnamefont {J.~M.~L.}\ \bibnamefont {Miller}}, \bibinfo {author} {\bibfnamefont {A.}~\bibnamefont {Ansari}}, \bibinfo {author} {\bibfnamefont {D.~B.}\ \bibnamefont {Heinz}}, \bibinfo {author} {\bibfnamefont {Y.}~\bibnamefont {Chen}}, \bibinfo {author} {\bibfnamefont {I.~B.}\ \bibnamefont {Flader}}, \bibinfo {author} {\bibfnamefont {D.~D.}\ \bibnamefont {Shin}}, \bibinfo {author} {\bibfnamefont {L.~G.}\ \bibnamefont {Villanueva}},\ and\ \bibinfo {author} {\bibfnamefont {T.~W.}\ \bibnamefont {Kenny}},\ }\bibfield  {title} {\bibinfo {title} {{Effective quality factor tuning mechanisms in micromechanical resonators}},\ }\href {https://doi.org/10.1063/1.5027850} {\bibfield  {journal} {\bibinfo  {journal} {Appl. Phys. Rev.}\ }\textbf {\bibinfo {volume} {5}},\ \bibinfo {pages} {4} (\bibinfo {year} {2018})}\BibitemShut {NoStop}%
\bibitem [{\citenamefont {Clark}\ \emph {et~al.}(2024)\citenamefont {Clark}, \citenamefont {Bernard}, \citenamefont {Ma}, \citenamefont {Dumont},\ and\ \citenamefont {Sankey}}]{Clark2024Nov}%
  \BibitemOpen
  \bibfield  {author} {\bibinfo {author} {\bibfnamefont {T.~J.}\ \bibnamefont {Clark}}, \bibinfo {author} {\bibfnamefont {S.}~\bibnamefont {Bernard}}, \bibinfo {author} {\bibfnamefont {J.}~\bibnamefont {Ma}}, \bibinfo {author} {\bibfnamefont {V.}~\bibnamefont {Dumont}},\ and\ \bibinfo {author} {\bibfnamefont {J.~C.}\ \bibnamefont {Sankey}},\ }\bibfield  {title} {\bibinfo {title} {{Optically Defined Phononic Crystal Defect}},\ }\href {https://doi.org/10.1103/PhysRevLett.133.226904} {\bibfield  {journal} {\bibinfo  {journal} {Phys. Rev. Lett.}\ }\textbf {\bibinfo {volume} {133}},\ \bibinfo {pages} {226904} (\bibinfo {year} {2024})}\BibitemShut {NoStop}%
\bibitem [{\citenamefont {Evans}\ \emph {et~al.}(2008)\citenamefont {Evans}, \citenamefont {Ballmer}, \citenamefont {Fejer}, \citenamefont {Fritschel}, \citenamefont {Harry},\ and\ \citenamefont {Ogin}}]{Evans2008Nov}%
  \BibitemOpen
  \bibfield  {author} {\bibinfo {author} {\bibfnamefont {M.}~\bibnamefont {Evans}}, \bibinfo {author} {\bibfnamefont {S.}~\bibnamefont {Ballmer}}, \bibinfo {author} {\bibfnamefont {M.}~\bibnamefont {Fejer}}, \bibinfo {author} {\bibfnamefont {P.}~\bibnamefont {Fritschel}}, \bibinfo {author} {\bibfnamefont {G.}~\bibnamefont {Harry}},\ and\ \bibinfo {author} {\bibfnamefont {G.}~\bibnamefont {Ogin}},\ }\bibfield  {title} {\bibinfo {title} {{Thermo-optic noise in coated mirrors for high-precision optical measurements}},\ }\href {https://doi.org/10.1103/PhysRevD.78.102003} {\bibfield  {journal} {\bibinfo  {journal} {Phys. Rev. D}\ }\textbf {\bibinfo {volume} {78}},\ \bibinfo {pages} {102003} (\bibinfo {year} {2008})}\BibitemShut {NoStop}%
\bibitem [{\citenamefont {Cerdonio}\ \emph {et~al.}(2001)\citenamefont {Cerdonio}, \citenamefont {Conti}, \citenamefont {Heidmann},\ and\ \citenamefont {Pinard}}]{cerdonio2001thermoelastic}%
  \BibitemOpen
  \bibfield  {author} {\bibinfo {author} {\bibfnamefont {M.}~\bibnamefont {Cerdonio}}, \bibinfo {author} {\bibfnamefont {L.}~\bibnamefont {Conti}}, \bibinfo {author} {\bibfnamefont {A.}~\bibnamefont {Heidmann}},\ and\ \bibinfo {author} {\bibfnamefont {M.}~\bibnamefont {Pinard}},\ }\bibfield  {title} {\bibinfo {title} {{Thermoelastic effects at low temperatures and quantum limits in displacement measurements}},\ }\href {https://doi.org/10.1103/PhysRevD.63.082003} {\bibfield  {journal} {\bibinfo  {journal} {Phys. Rev. D}\ }\textbf {\bibinfo {volume} {63}},\ \bibinfo {pages} {082003} (\bibinfo {year} {2001})}\BibitemShut {NoStop}%
\bibitem [{\citenamefont {De~Rosa}\ \emph {et~al.}(2002)\citenamefont {De~Rosa}, \citenamefont {Conti}, \citenamefont {Cerdonio}, \citenamefont {Pinard},\ and\ \citenamefont {Marin}}]{deRosa2002experimental}%
  \BibitemOpen
  \bibfield  {author} {\bibinfo {author} {\bibfnamefont {M.}~\bibnamefont {De~Rosa}}, \bibinfo {author} {\bibfnamefont {L.}~\bibnamefont {Conti}}, \bibinfo {author} {\bibfnamefont {M.}~\bibnamefont {Cerdonio}}, \bibinfo {author} {\bibfnamefont {M.}~\bibnamefont {Pinard}},\ and\ \bibinfo {author} {\bibfnamefont {F.}~\bibnamefont {Marin}},\ }\bibfield  {title} {\bibinfo {title} {{Experimental Measurement of the Dynamic Photothermal Effect in Fabry-Perot Cavities for Gravitational Wave Detectors}},\ }\href {https://doi.org/10.1103/PhysRevLett.89.237402} {\bibfield  {journal} {\bibinfo  {journal} {Phys. Rev. Lett.}\ }\textbf {\bibinfo {volume} {89}},\ \bibinfo {pages} {237402} (\bibinfo {year} {2002})}\BibitemShut {NoStop}%
\bibitem [{\citenamefont {Leviton}\ and\ \citenamefont {Frey}(2006)}]{Leviton2006Jul}%
  \BibitemOpen
  \bibfield  {author} {\bibinfo {author} {\bibfnamefont {D.~B.}\ \bibnamefont {Leviton}}\ and\ \bibinfo {author} {\bibfnamefont {B.~J.}\ \bibnamefont {Frey}},\ }\bibfield  {title} {\bibinfo {title} {{Temperature-dependent absolute refractive index measurements of synthetic fused silica}},\ }\bibfield  {booktitle} {\emph {\bibinfo {booktitle} {{Proceedings Volume 6273, Optomechanical Technologies for Astronomy}}},\ }\href {https://doi.org/10.1117/12.672853} {\bibfield  {journal} {\bibinfo  {journal} {Optomechanical Technologies for Astronomy}\ }\textbf {\bibinfo {volume} {6273}},\ \bibinfo {pages} {800} (\bibinfo {year} {2006})}\BibitemShut {NoStop}%
\bibitem [{\citenamefont {Fejer}\ \emph {et~al.}(2004)\citenamefont {Fejer}, \citenamefont {Rowan}, \citenamefont {Cagnoli}, \citenamefont {Crooks}, \citenamefont {Gretarsson}, \citenamefont {Harry}, \citenamefont {Hough}, \citenamefont {Penn}, \citenamefont {Sneddon},\ and\ \citenamefont {Vyatchanin}}]{Fejer2004Oct}%
  \BibitemOpen
  \bibfield  {author} {\bibinfo {author} {\bibfnamefont {M.~M.}\ \bibnamefont {Fejer}}, \bibinfo {author} {\bibfnamefont {S.}~\bibnamefont {Rowan}}, \bibinfo {author} {\bibfnamefont {G.}~\bibnamefont {Cagnoli}}, \bibinfo {author} {\bibfnamefont {D.~R.~M.}\ \bibnamefont {Crooks}}, \bibinfo {author} {\bibfnamefont {A.}~\bibnamefont {Gretarsson}}, \bibinfo {author} {\bibfnamefont {G.~M.}\ \bibnamefont {Harry}}, \bibinfo {author} {\bibfnamefont {J.}~\bibnamefont {Hough}}, \bibinfo {author} {\bibfnamefont {S.~D.}\ \bibnamefont {Penn}}, \bibinfo {author} {\bibfnamefont {P.~H.}\ \bibnamefont {Sneddon}},\ and\ \bibinfo {author} {\bibfnamefont {S.~P.}\ \bibnamefont {Vyatchanin}},\ }\bibfield  {title} {\bibinfo {title} {{Thermoelastic dissipation in inhomogeneous media: loss measurements and displacement noise in coated test masses for interferometric gravitational wave detectors}},\ }\href {https://doi.org/10.1103/PhysRevD.70.082003} {\bibfield  {journal} {\bibinfo  {journal} {Phys. Rev. D}\ }\textbf {\bibinfo
  {volume} {70}},\ \bibinfo {pages} {082003} (\bibinfo {year} {2004})}\BibitemShut {NoStop}%
\end{thebibliography}

\clearpage
\end{document}